\documentclass[twocolumn,showpacs,preprintnumbers,amsmath,amssymb]{revtex4}

\usepackage{graphicx}
\usepackage{dcolumn}
\usepackage{bm}

\begin{document}

\title{How to project a bipartite network?}
\author{Tao Zhou$^{1,2}$}
\email{zhutou@ustc.edu}
\author{Jie Ren$^1$}
\author{Mat\'u\v{s} Medo$^1$}
\author{Yi-Cheng Zhang$^{1,3}$}
\email{yi-cheng.zhang@unifr.ch}
\affiliation{%
$^1$Department of Physics, University of Fribourg, Chemin du Muse
3, CH-1700 Fribourg, Switzerland \\
$^2$Department of Modern Physics and Nonlinear Science Center,
University of Science and Technology of China, Hefei Anhui,
230026, PR China \\
$^3$Information Economy and Internet Research Laboratory,
University of Electronic Science and Technology of China, Chengdu
Sichuan, 610054, PR China
}%

\date{\today}

\begin{abstract}
The one-mode projecting is extensively used to compress the
bipartite networks. Since the one-mode projection is always less
informative than the bipartite representation, a proper weighting
method is required to better retain the original information. In
this article, inspired by the network-based resource-allocation
dynamics, we raise a weighting method, which can be directly
applied in extracting the hidden information of networks, with
remarkably better performance than the widely used global ranking
method as well as collaborative filtering. This work not only
provides a creditable method in compressing bipartite networks,
but also highlights a possible way for the better solution of a
long-standing challenge in modern information science: How to do
personal recommendation?
\end{abstract}


\maketitle

\section{Introduction}
The last few years have witnessed a tremendous activity devoted to
the understanding of complex networks
\cite{Amaral2000,Strogatz2001,Albert2002,Dorogovtsev2002,Newman2003,Boccaletti2006,Costa2007}.
A particular class of networks is the \emph{bipartite networks},
whose nodes are divided into two sets, $X$ and $Y$, and only the
connection between two nodes in different sets is allowed (as
illustrated in Fig. 1a). Many systems are naturally modeled as
bipartite networks \cite{Holme2003}: Human sexual network
\cite{Liljeros2001} is consisted of men and women, metabolic
network \cite{Jeong2000} is consisted of chemical substances and
chemical reactions, etc. Two kinds of bipartite networks should be
paid more attention for their particular significance in social,
economic and information systems. One is the so-called
\emph{collaboration network}, which is generally defined as a
networks of actors connected by a common collaboration act
\cite{Wasserman1994,Scott2000}. Examples are numerous, including
scientists connected by coauthoring a scientific paper
\cite{Newman2001a,Newman2001b}, movie actors connected by
costarring the same movie \cite{Watts1998,Amaral2000}, and so on.
Moreover, the concept of collaboration network is not necessarily
restricted within social systems (see, for example, recent reports
on technological collaboration of software \cite{Myers2003} and
urban traffic systems \cite{Zhang2006}). Although the
collaboration network is usually displayed by the one-mode
projection on actors (see later the definition), its fully
representation is a bipartite network. The other one is named
\emph{opinion network} \cite{Maslov2001,Blattner2007}, where each
node in the \emph{user-set} is connected with its collected
objects in the \emph{object-set}. For example, listeners are
connected with the music groups they collected from music-sharing
library (e.g. \emph{audioscrobbler.com})
\cite{Lambiotte2005,Cano2006}, web-users are connected with the
webs they collected in a bookmark site (e.g. \emph{delicious})
\cite{Cattuto2007}, customers are connected with the books they
bought (e.g. \emph{Amazon.com}) \cite{Linden2003,Yammine2004}.

Recently, a large amount of attention is addressed to analyzing
\cite{Holme2003,Lambiotte2005,Lambiotte2005b,Lind2005,Estrada2005}
and modeling \cite{Ramasco2004,Ohkubo2005,Peltomaki2006} bipartite
network. However, for the convenience of directly showing the
relations among a particular set of nodes, the bipartite network
is usually compressed by one-mode projecting. The one-mode
projection onto $X$ ($X$-projection for short) means a network
containing only $X$-nodes, where two $X$-nodes are connected when
they have at least one common neighboring $Y$-node. Fig. 1b and
Fig. 1c show the resulting networks of $X$-projection and
$Y$-projection, respectively. The simplest way is to project the
bipartite network onto an unweighted network
\cite{Newman2001a,Newman2001b,Grossman1995,Barabasi2002,Zhou2007},
without taking into account of the frequency that a collaboration
has been repeated. Although some topological properties can be
qualitatively obtained from this unweighted version, the loss of
information is obvious. For example, if two listeners has
collected more than 100 music groups each (it is a typical number
of collections, like in \emph{audioscrobbler.com}, the average
number of collected music groups per listener is 140
\cite{Lambiotte2005}), and only one music group is selected by
both listeners, one may conclude that those two listeners probably
have different music taste. On the contrary, if nearly 100 music
groups belong to the overlap, those two listeners are likely to
have very similar habits. However, in the unweighted
listener-projection, this two cases have exactly the same graph
representation.

Since the one-mode projection is always less informative than the
original bipartite network, to better reflect structure of the
network, one has to use the bipartite graph to quantify the
weights in the projection graph. A straightforward way is to
weight an edge directly by the number of times the corresponding
partnership repeated \cite{Ramosco2006,Li2007}. This simple rule
is used to obtain the weights in Fig. 1b and Fig. 1c for
$X$-projection and $Y$-projection, respectively. This weighted
network is much more informative than the unweighted one, and can
be analyzed by standard techniques for unweighted graphs since its
weights are all integers \cite{Newman2004a}. However, this method
is also quantitatively biased. Li \emph{et al.} \cite{Li2005}
empirically studied the scientific collaboration networks, and
pointed out that the impact of one additional collaboration paper
should depend on the original weight between the two scientists.
For example, one more co-authorized paper for the two authors
having only co-authorized one paper before should have higher
impact than for the two authors having already co-authorized 100
papers. This saturation effect can be taken into account by
introducing a hyperbolic tangent function onto the simple count of
collaborated times \cite{Li2005}. As stated by Newman that two
scientists whose names appear on a paper together with many other
coauthors know one another less well on average than two who were
the sole authors of a paper \cite{Newman2001b}, to consider this
effect, he introduced the factor $1/(n-1)$ to weaken the
contribution of collaborations involving many participants
\cite{Newman2001c,Newman2004b}, where $n$ is the number of
participants (e.g. the number of authors of a paper).

\begin{figure}
\scalebox{0.8}[0.8]{\includegraphics{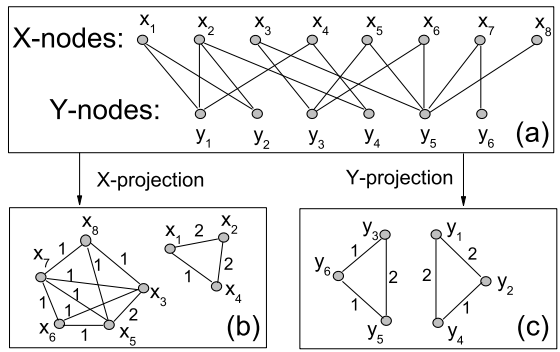}}
\caption{Illustration of a bipartite network (a), as well as its
$X$-projection (b) and $Y$-projection (c). The edge-weight in (b)
and (c) is set as the number of common neighbors in $Y$ and $X$,
respectively.}
\end{figure}

How to weight the edges is the key question of the one-mode
projections and their use. However, we lack a systematic
exploration of this problem, and no solid base of any weighting
methods have been reported thus far. For example, one may ask the
physical reason why using the hyperbolic tangent function to
address the saturation effect \cite{Li2005} rather than other
infinite possible candidates. In addition, for simplicity, the
weighted adjacent matrix $\{w_{ij}\}$ is always set to be
symmetrical, that is, $w_{ij}=w_{ji}$. However, as in scientific
collaboration networks, different authors may assign different
weights to the same co-authorized paper, and it is probably the
case that the author having less publications may give a higher
weight, vice versa. Therefore, a more natural weighting method may
be not symmetrical. Another blemish in the prior methods is that
the information contained by the edge whose adjacent $X$-node
($Y$-node) is of degree one will be lost in $Y$-projection
($X$-projection). This information loss may be serious in some
real opinion networks. For example, in the user-web network of
\emph{delicious} (http://del.icio.us), a remarkable fraction of
webs have been collected only once, as well as a remarkable
fraction of users have collected only one web. Therefore, both the
user-projection and web-projection will squander a lot of
information. Since more than half publications in
\emph{Mathematical Reviews} have only one author
\cite{Grossman1995}, the situation is even worse in mathematical
collaboration network.

In this article, we propose a weighting method, with asymmetrical
weights (i.e., $w_{ij}\neq w_{ji}$) and allowed self-connection
(i.e., $w_{ii}>0$). This method can be directly applied as a
personal recommendation algorithm, which performs remarkably
better than the widely used \emph{global ranking method} (GRM) and
\emph{collaborative filtering} (CF).

\section{Method}
Without loss of generality, we discuss how to determine the
edge-weight in $X$-projection, where the weight $w_{ij}$ can be
considered as the importance of node $i$ in $j$'s sense, and it is
generally not equal to $w_{ji}$. For example, in the
book-projection of a customer-book opinion network, the weight
$w_{ij}$ between two books $i$ and $j$ contributes to the strength
of book $i$ recommendation to a customer provided he has brought
book $j$. In the scientific collaboration network, $w_{ij}$
reflects how likely is $j$ to choose $i$ as a contributor for a
new research project. More generally, we assume a certain amount
of a resource (e.g. recommendation power, research fund, etc.) is
associated with each $X$-node, and the weight $w_{ij}$ represents
the proportion of the resource $j$ would like to distribute to
$i$.

\begin{figure}
\scalebox{1}[1.2]{\includegraphics{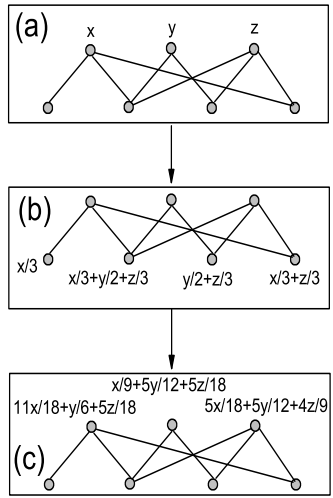}} \caption{Illustration
of the resource-allocation process in bipartite network. The upper
three are $X$-nodes, and the lower four are $Y$-nodes. The whole
process consists of two steps: First, the resource flows from $X$
to $Y$ (a$\rightarrow$b), and then returns to $X$
(b$\rightarrow$c). Different from the prior network-based
resource-allocation dynamics \cite{Ou2007}, the resource here can
only flow from one node-set to another node-set, without
consideration of asymptotical stable flow among one node-set.}
\end{figure}

To derive the analytical expression of $w_{ij}$, we go back to the
bipartite representation. Since the bipartite network itself is
unweighted, the resource in an arbitrary $X$-node should be
equally distributed to its neighbors in $Y$. Analogously, the
resource in any $Y$-node should be equally distributed to its
$X$-neighbors. As shown in Fig. 2a, the three $X$-nodes are
initially assigned weights $x$, $y$ and $z$. The
resource-allocation process consists of two steps; first from $X$
to $Y$, then back to $X$. The amount of resource after each step
is marked in Fig. 2b and Fig. 2c, respectively. Merging these two
steps into one, the final resource located in those three
$X$-nodes, denoted by $x'$, $y'$ and $z'$, can be obtained as:
\begin{equation}
 \left(
     \begin{array}{c}
        x' \\
        y' \\
        z' \\
     \end{array}
     \right)
 =\left(
    \begin{array}{ccc}
        11/18 & 1/6 & 5/18 \\
        1/9 & 5/12 & 5/12 \\
        5/18 & 5/12 & 4/9
    \end{array}
    \right)
 \left(
     \begin{array}{c}
        x \\
        y \\
        z \\
     \end{array}
     \right).
\end{equation}
Note that, this $3 \times 3$ matrix are column normalized, and the
element in the $i$th row and $j$th column represents the fraction
of resource the $j$th $X$-node transferred to the $i$th $X$-node.
According to the above description, this matrix is the very
weighted adjacent matrix we want.

Now, consider a general bipartite network $G(X,Y,E)$, where $E$ is
the set of edges. The nodes in $X$ and $Y$ are denoted by
$x_1,x_2,\cdots,x_n$ and $y_1,y_2,\cdots,y_m$, respectively. The
initial resource located on the $i$th $X$-node is $f(x_i)\geq 0$.
After the first step, all the resource in $X$ flows to $Y$, and
the resource located on the $l$th $Y$-node reads,
\begin{equation}
f(y_l)=\sum^n_{i=1}\frac{a_{il}f(x_i)}{k(x_i)},
\end{equation}
where $k(x_i)$ is the degree of $x_i$, and $a_{il}$ is an $n
\times m$ adjacent matrix as
\begin{equation}
a_{il}=\left\{
    \begin{array}{cc}
       1,  &  x_iy_l\in E, \\
       0,  &  \texttt{otherwise}. \\
    \end{array}
    \right.
\end{equation}
In the next step, all the resource flows back to $X$, and the
final resource located on $x_i$ reads,
\begin{equation}
f'(x_i)=\sum^m_{l=1}a_{il}f(y_l)/k(y_l)=\sum^m_{l=1}\frac{a_{il}}{k(y_l)}\sum^n_{j=1}\frac{a_{jl}f(x_j)}{k(x_j)}.
\end{equation}
This can be rewritten as
\begin{equation}
f'(x_i)=\sum^n_{j=1}w_{ij}f(x_j),
\end{equation}
where
\begin{equation}
w_{ij}=\frac{1}{k(x_j)}\sum^m_{l=1}\frac{a_{il}a_{jl}}{k(y_l)},
\end{equation}
which sums the contribution from all 2-step paths between $x_i$
and $x_j$. The matrix $W=\{w_{ij}\}_{n\times n}$ represents the
weighted $X$-projection we were looking for. The
resource-allocation process can be written in the matrix form as
$\overrightarrow{f'}=W\overrightarrow{f}$.

It is worthwhile to emphasize the particular characters of this
weighting method. For convenience, we take the scientific
collaboration network as an example, but our statements are not
restricted to the collaboration networks. Firstly, the weighted
matrix is not symmetrical as
\begin{equation}
\frac{w_{ij}}{k(x_j)}=\frac{w_{ji}}{k(x_i)}.
\end{equation}
This is in accordance with our daily experience - the weight of a
single collaboration paper is relatively small if the scientist
has already published many papers (i.e., he has large degree),
vice versa. Secondly, the diagonal elements in $W$ are nonzero,
thus the information contained by the connections incident to
one-degree $Y$-node will not be lost. Actually, the diagonal
element is the maximal element in each column. Only if all $x_i$'s
$Y$-neighbors belongs to $x_j$'s neighbors set, $w_{ii}=w_{ji}$.
It is usually found in scientific collaboration networks, since
some students coauthorize every paper with their supervisors.
Therefore, the ratio $w_{ji}/w_{ii}\leq 1$ can be considered as
$x_i$'s researching independence to $x_j$, the smaller the ratio,
the more independent the researcher is, vice versa. The
independence of $x_i$ can be approximately measured as
\begin{equation}
I_i=\sum_j\left(\frac{w_{ji}}{w_{ii}}\right)^2.
\end{equation}
Generally, the author who often publishes papers solely, or often
publishes many papers with different coauthors is more
independent. Note that, introducing the measure $I_i$ here is just
to show an example how to use the information contained by
self-weight $w_{ii}$, without any comments whether to be more
independent is better, or contrary.

\section{Personal recommendation}
The exponential growth of the Internet \cite{Faloutsos1999} and
World-Wide-Web \cite{Broder2000} confronts people with an
information overload: They are facing too many data and sources to
be able to find out those most relevant for him. One landmark for
information filtering is the use of search engines
\cite{Kleinberg1999}, however, it can not solve this
\emph{overload problem} since it does not take into account of
personalization thus returns the same results for people with far
different habits. So, if user's habits are different from the
mainstream, it is hard for him to find out what he likes in the
countless searching results. Thus far, the most potential way to
efficiently filter out the information overload is to recommend
personally. That is to say, using the personal information of a
user (i.e., the historical track of this user's activities) to
uncover his habits and to consider them in the recommendation. For
instances, Amazon.com uses one's purchase history to provide
individual suggestions. If you have bought a textbook on
statistical physics, Amazon may recommend you some other
statistical physics books. Based on the well-developed \emph{Web
2.0} technology \cite{Alexander2006}, the recommendation systems
are frequently used in web-based movie-sharing (music-sharing,
book-sharing, etc.) systems, web-based selling systems, bookmark
web-sites, and so on. Motivated by the significance in economy and
society, recently, the design of an efficient recommendation
algorithm becomes a joint focus from marketing practice
\cite{Ansari2000,Ying2006} to mathematical analysis
\cite{Kumar2001}, from engineering science
\cite{Belkin2000,Montaner2003,Herlocker2004} to physics community
\cite{Laureti2006,Yu2006,Walter2006}.

\begin{figure}
\scalebox{0.8}[0.8]{\includegraphics{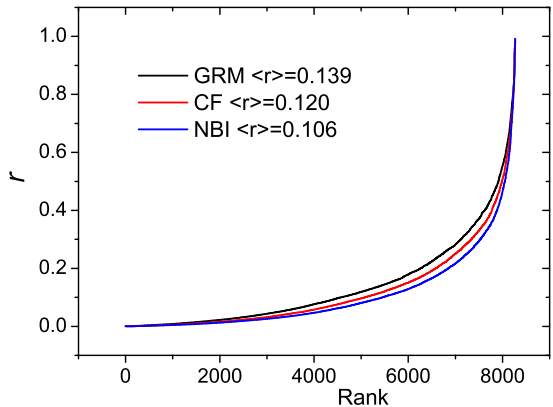}} \caption{(color
online) The predicted position of each entry in the probe ranked
in the ascending order. The black, red and blue curves, from top
to bottom, represent the cases of GRM, CF and NBI, respectively.
The mean values are top 13.9\% (GRM), top 12.0\% (CF) and top
10.6\% (NBI).}
\end{figure}

Basically, a recommendation system consists of users and objects,
and each user has collected some objects. Denote the object-set as
$O=\{o_1,o_2,\cdots,o_n\}$ and user-set as
$U=\{u_1,u_2,\cdots,u_m\}$. If users are only allowed to collect
objects (they do not rate them), the recommendation system can be
fully described by an $n\times m$ adjacent matrix $\{a_{ij}\}$,
where $a_{ij}=1$ if $u_j$ has already collected $o_i$, and
$a_{ij}=0$ otherwise. A reasonable assumption is that the objects
you have collected are what you like, and a recommendation
algorithm aims at predicting your personal opinions (to what
extent you like or hate them) on those objects you have not yet
collected. A more complicated case is the voting system
\cite{Konstan1997,Goldberg2001}, where each user can give ratings
to objects (e.g., in the \emph{Yahoo Music}, the users can vote
each song with 5 discrete ratings representing \emph{Never play
again}, \emph{It is ok}, \emph{Like it}, \emph{Love it}, and
\emph{Can't get enough}), and the recommendation algorithm
concentrates on estimating unknown ratings for objects. These two
problems are closely related, however, in this article, we focus
on the former case.

Denote $k(o_i)=\sum^m_{j=1}a_{ij}$ the degree of object $o_i$. The
\emph{global ranking method} (GRM) sorts all the objects in the
descending order of degree and recommends those with highest
degrees. Although the lack of personalization leads to an
unsatisfying performance of GRM (see numerical comparison in the
next section), it is widely used since it is simple and spares
computational resources. For example, the well-known \emph{Yahoo
Top 100 MTVs}, \emph{Amazon List of Top Sellers}, as well as the
board of most downloaded articles in many scientific journals, can
be all considered as results of GRM.

Thus far, the widest applied personal recommendation algorithm is
\emph{collaborative filtering} (CF)
\cite{Herlocker2004,Konstan1997}, based on a similarity measure
between users. Consequently, the prediction for a particular user
is made mainly using the similar users. The similarity between
users $u_i$ and $u_j$ can be measured in the Pearson-like form
\begin{equation}
s_{ij}=\frac{\sum^n_{l=1}a_{li}a_{lj}}{\texttt{min}\{k(u_i),k(u_j)\}},
\end{equation}
where $k(u_i)=\sum^n_{l=1}a_{li}$ is the degree of user $u_i$. For
any user-object pair $u_i-o_j$, if $u_i$ has not yet collected
$o_j$ (i.e., $a_{ji}=0$), by CF, the predicted score, $v_{ij}$ (to
what extent $u_i$ likes $o_j$), is given as
\begin{equation}
v_{ij}=\frac{\sum^m_{l=1,l\neq i}s_{li}a_{jl}}{\sum^m_{l=1,l\neq
i}s_{li}}.
\end{equation}
Two factors give rise to a high value of $v_{ij}$. Firstly, if the
degree of $o_j$ is larger, it will, generally, have more nonzero
items in the numerator of Eq. (10). Secondly, if $o_j$ is
frequently collected by users very similar to $u_i$, the
corresponding items will be significant. The former pays respect
to the global information, and the latter reflects the
personalization. For any user $u_i$, all the nonzero $v_{ij}$ with
$a_{ji}=0$ are sorted in descending order, and those objects in
the top are recommended.

We propose a recommendation algorithm, which is a direct
application of the weighting method for bipartite networks
presented above. The layout is simple: first compress the
bipartite user-object network by object-projection, the resulting
weighted network we label $G$. Then, for a given user $u_i$, put
some resource on those objects already been collected by $u_i$.
For simplicity, we set the initial resource located on each node
of $G$ as
\begin{equation}
f(o_j)=a_{ji}.
\end{equation}
That is to say, if the object $o_j$ has been collected by $u_i$,
then its initial resource is unit, otherwise it is zero. Note
that, the initial configuration, which captures personal
preferences, is different for different users. The initial
resource can be understood as giving a unit recommending capacity
to each collected object. According to the weighted
resource-allocation process discussed in the prior section, the
final resource, denoted by the vector $\overrightarrow{f'}$, is
$\overrightarrow{f'}=W\overrightarrow{f}$. Thus components of $f'$
are
\begin{equation}
f'(o_j)=\sum^n_{l=1}w_{jl}f(o_l)=\sum^n_{l=1}w_{jl}a_{li}.
\end{equation}
For any user $u_i$, all his uncollected objects $o_j$ ($1\leq j
\leq n$, $a_{ji}=0$) are sorted in the descending order of
$f'(o_j)$, and those objects with highest value of final resource
are recommended. We call this method \emph{network-based
inference} (NBI), since it is based on the weighted network $G$.
Note that, the calculation of Eq. (12) should be repeated $m$
times, since the initial configurations are different for
different users.

\section{Numerical results}
We use a benchmark data-set, namely \emph{MovieLens}, to judge the
performance of described algorithms. The MovieLens data is
downloaded from the web-site of \emph{GroupLens Research}
(http://www.grouplens.org). The data consists 1682 movies
(objects) and 943 users. Actually, MovieLens is a rating system,
where each user votes movies in five discrete ratings 1-5. Hence
we applied the coarse-graining method similar to what is used in
Ref. \cite{Blattner2007}: A movie has been collected by a user iff
the giving rating is at least 3. The original data contains $10^5$
ratings, 85.25\% of which are $\geq 3$, thus the user-movie
bipartite network after the coarse gaining contains 85250 edges.
To test the recommendation algorithms, the data set (i.e., 85250
edges) is randomly divided into two parts: The training set
contains 90\% of the data, and the remaining 10\% of data
constitutes the probe. The training set is treated as known
information, while no information in probe set is allowed to be
used for prediction.

\begin{figure}
\scalebox{0.8}[0.8]{\includegraphics{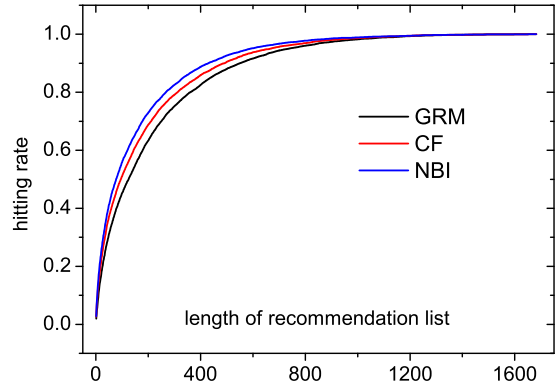}} \caption{The hitting
rate as a function of the length of recommendation list. The
black, red and blue curves, from bottom to top, represent the
cases of GRM, CF and NBI, respectively.}
\end{figure}

All three algorithms, GRM, CF and NBI, can provide each user an
ordered queue of all its uncollected movies. For an arbitrary user
$u_i$, if the edge $u_i-o_j$ is in the probe set (according to the
training set, $o_j$ is an uncollected movie for $u_i$), we measure
the position of $o_j$ in the ordered queue. For example, if there
are 1500 uncollected movies for $u_i$, and $o_j$ is the 30th from
the top, we say the position of $o_j$ is the top 30/1500, denoted
by $r_{ij}=0.02$. Since the probe entries are actually collected
by users, a good algorithm is expected to give high
recommendations to them, thus leading to small $r$. The mean value
of the position value, averaged over entries in the probe, are
0.139, 0.120 and 0.106 by GRM, CF and NBI, respectively. Fig. 3
reports the distribution of all the position values, which are
ranked from the top position ($r\rightarrow 0$) to the bottom
position ($r\rightarrow 1$). Clearly, NBI is the best method and
GRM performs worst.

\begin{table}
\renewcommand{\arraystretch}{1.3}
\caption{The hitting rates for some typical lengths of
recommendation list.}
\begin{tabular}{cccc}
Length & GRM & CF & NBI\\
\hline 10 & 10.3\% & 14.1\% & 16.2\% \\
\hline 20 & 16.9\%  & 21.6\% & 24.8\% \\
\hline 50 & 31.1\% & 37.0\% & 41.2\% \\
\hline 100 & 45.2\% & 51.0\% & 55.9\% \\
\hline
\end{tabular}
\end{table}

To make this work more relevant to the real-life recommendation
systems, we introduce a measure of algorithmic accuracy that
depends on the length of recommendation list. The recommendation
list for a user $u_i$, if of length $L$, contains $L$ highest
recommended movies resulting from the algorithm. For each incident
entry $u_i-o_j$ in the probe, if $o_j$ is in $u_i$'s
recommendation list, we say the entry $u_i-o_j$ is \emph{hit} by
the algorithm. The ratio of hit entries to the population is named
\emph{hitting rate}. For a given $L$, the algorithm with a higher
hitting rate is better, and vice versa. If $L$ is larger than the
total number of uncollected movies for a user, the recommendation
list is defined as the set of all his uncollected movies. Clearly,
the hitting rate is monotonously increasing with $L$, with the
upper bound 1 for sufficiently large $L$. In Fig. 4, we report the
hitting rate as a function of $L$ for different algorithms. In
accordance with Fig. 3, the accuracy of the algorithms is NBI $>$
CF $>$ GRM. The hitting rates for some typical lengths of
recommendation list are shown in Table I.

In a word, via the numerical calculation on a benchmark data set,
we have demonstrated that the NBI has remarkably better
performance than GRM and CF, which strongly guarantee the validity
of the present weighting method.

\section{Conclusion and Discussion}
Weighting of edges is the key problem in the construction of a
bipartite network projection. In this article we proposed a
weighting method based on a resource-allocation process. The
present method has two prominent features. First, the weighted
matrix is not symmetrical, and the node having larger degree in
the bipartite network generally assigns smaller weights to its
incident edges. Second, the diagonal element in the weighted
matrix is positive, which makes the weighted one-mode projection
more informative.

Furthermore, we proposed a personal recommendation algorithm based
on this weighting method, which performs much better than the
widest used global ranking method as well as the collaborative
filtering. Especially, this algorithm is tune-free (i.e., does not
depend on any control parameters), which is a big advantage for
potential users. The main goal of this article is to raise a new
weighting method, as well as provide a bridge from this method to
the recommendation systems. The presented recommendation algorithm
is just a rough framework, whose details have not been
exhaustively explored yet. For example, the setting of the initial
configuration may be oversimplified, a more complicated form, like
$f(o_j)=a_{ji}k^{\beta}(o_j)$, may lead to a better performance
than the presented one with $\beta=0$. One is also encouraged to
consider the asymptotical dynamics of the resource-allocation
process \cite{Ou2007}, which can eventually lead to some certain
iterative recommendation algorithms. Although such an algorithm
require much longer CPU time, it may give more accurate prediction
than the present algorithm.

If we denote $\langle k_u\rangle$ and $\langle k_o\rangle$ the
average degree of users and objects in the bipartite network, the
computational complexity of CF is $\mathbb{O}(m^2\langle
k_u\rangle+mn\langle k_o\rangle)$, where the first term accounts
for the calculation of similarity between users (see Eq. (9)), and
the second term accounts for the calculation of the predicted
score (see Eq. (10)). Substituting the equation $n\langle
k_o\rangle=m\langle k_u\rangle$, we are left with
$\mathbb{O}(m^2\langle k_u\rangle)$. The computational complexity
for NBI is $\mathbb{O}(m\langle k_u^2\rangle+mn\langle
k_u\rangle)$ with two terms accounting for the calculation of the
weighted matrix and the final resource distribution, respectively.
Here $\langle k_u^2\rangle$ is the second moment of the users'
degree distribution in the bipartite network. Clearly, $\langle
k_u^2\rangle<n\langle k_u\rangle$, thus the resulting form is
$\mathbb{O}(mn\langle k_u\rangle)$. Note that the number of users
is usually much larger than the number of objects in many
recommendation systems. For instance, the \emph{EachMovie} dataset
provided by the \emph{Compaq} company contains $m=72916$ users and
$n=1628$ movies, and the \emph{Netflix} company provides nearly 20
thousands online movies for million users. It is also the case of
music-sharing systems and online bookstores, the number of
registered users is more than one magnitude larger than that of
the available objects (e.g., music groups, books, etc.).
Therefore, NBI runs much fast than CF. In addition, NBI requires
$n^2$ memory to store the weighted matrix $\{w_{ij}\}$, while CF
requires $m^2$ memory to store the similarity matrix $\{s_{ij}\}$.
Hence, NBI is able to beat CF in all the three criterions of
recommendation algorithm: \emph{accuracy}, \emph{time} and
\emph{space}. However, in some recommendation systems, as in
bookmark sharing websites, the number of objects (e.g. webpages)
is much larger than the number of users, thus CF may be more
practicable.

\begin{acknowledgments}
The authors thank to Sang Hoon Lee for his comments and suggestions.
This work is partially supported by Swiss National Science
Foundation (project 205120-113842). We acknowledge SBF (Switzerland)
for financial support through project C05.0148 (Physics of Risk),
TZhou acknowledges the NNSFC under Grant No. 10635040.
\end{acknowledgments}

\end{document}